\def\ch{}
\documentclass[cm]{aa}
\usepackage{graphics}
\usepackage{longtable}
\usepackage{supertabular}
\usepackage{astron}
\usepackage{graphicx,subfigure}
\newenvironment{longtable*}
               {\@dblfloat{longtable}}
               {\end@dblfloat}
\usepackage{epsfig}
\newcommand{\beq}{\begin{equation}}
\newcommand{\beqa}{\begin{eqnarray}}
\newcommand{\eeq}{\end{equation}}
\newcommand{\eeqa}{\end{eqnarray}}
\begin{document}
   \thesaurus{06         
              (08.22.3;  
	       03.13.2;  
               08.16.4;  
	       03.20.4;  
               11.13.1;  
	       08.05.3)} 
                                            
\title{AGAPEROS: Searching for variable stars in the LMC Bar with the
Pixel Method\thanks{This work is based on data collected by the EROS
collaboration.}}  \subtitle{I. Detection, astrometry and
cross-identification\thanks{Tables 2 is also available
in electronic form at the CDS via anonymous ftp to cdsarc.u-strasbg.fr
(130.79.128.5) or via http://cdsweb.u-strasbg.fr/Abstract.html}}

\author{A.-L. Melchior\inst{1,2} \and S.M.G. Hughes\inst{3} \and
J. Guibert\inst{4}}

\offprints{Anne-Laure Melchior}
\mail{A.L.Melchior@obspm.fr}

\institute{Astronomy Unit, Queen Mary and Westfield College, Mile End Road,
London E1\,4NS, UK
\and DEMIRM (UMR~8540), Observatoire de Paris,  61, Avenue de l'Observatoire, 
75\,014 Paris Cedex, France 
\and Institute of Astronomy, University of Cambridge, Madingley Road, 
Cambridge CB3\,0EZ, UK 
\and Centre d'Analyse des Images de l'INSU (UMR~8633), Observatoire de
Paris, 61 avenue de l'Observatoire, 75\,014 Paris, France
}

\date{Received / Accepted}
\titlerunning{Searching for variable stars in the LMC...}
\authorrunning{A.-L. Melchior et al.}
\maketitle

\begin{abstract} 
{\ch We extend the work developed in previous papers on microlensing
with a selection of variable stars. We use the Pixel Method to select
variable stars on a set of 2.5$\times 10^{6}$ pixel light curves in
the LMC Bar presented elsewhere. The previous treatment was done in
order to optimise the detection of long timescale variations (larger
than a few days) and we further optimise our analysis for the
selection of Long Timescale and Long Period Variables (LT$\&$LPV). We
choose to perform a selection of variable objects as comprehensive as
possible, independent of periodicity and of their position on the
colour magnitude diagram.  We detail the different thresholds
successively applied to the light curves, which allow to produce a
catalogue of 632 variable objects. We present a table with the
coordinate of each variable, its EROS magnitudes at one epoch and an
indicator of blending in both colours, together with a finding chart.

A cross-correlation with various catalogues shows that 90$\%$ of those
variable objects were undetected before, thus enlarging the sample of
{\ch LT$\&$LPV} previously known in this area by a factor of 10. Due to the
limitations of both the Pixel Method and the data set,} additional
data {\ch -- namely a longer baseline and near infrared photometry --
} are required to further characterise these variable stars, as will
be addressed in subsequent papers. 

\end{abstract}	

\section{Introduction}
The microlensing searches, motivated by the study of dark matter in
galaxies and its possible fraction of compact objects (hereafter
Machos), have collected some unprecedented data\-bases of images of
neighbouring galaxies, and in particular of the Large Magellanic Cloud
(LMC). The microlensing candidates are compatible with large Macho
masses, and exhibit long timescale variations, typically 10-200 days
(Alcock et al.\ 1997abc, Renault et al.\
1997\nocite{Alcock:1997a,Renault:1997}, Pa\-lan\-que-Delabrouille
et~al.  1998\nocite{Alcock:1997b,Palanque-Delabrouille:1998}, Alard et
al.\ 1997, Udalski et al 1994, Ansari et
al. 1999\nocite{Alcock:1997d,Alard:1997,Udalski:1994b,Ansari:1999}).
{\ch Possible contamination by variable stars is often suggested
(e.g. della Valle $\&$ Livio 1996\nocite{dellaValle:1996}), but no
systematic searches for variable objects have been undertaken so far
on the microlensing databases. Whereas particular variables, such as
RR Lyrae and cepheids are relatively well understood, little
statistical information is known about the {\ch Long Timescale and}
Long Period Variable stars (hereafter {\ch LT$\&$LPV}), with
timescales/periods in the range of $\sim$100 to $\sim$800 days, which
are an important stage of stellar evolution. They are easily rejected
for these microlensing searches either with a cut on marginal stellar
populations identified on the colour magnitude diagram (hereafter CMD)
or by their periodicity. Aperiodic signals which do not satisfy these
simple criteria are however more difficult to discriminate and to
select.}

In this paper, we perform a selection of variable objects on pixel
light curves covering a 0.25 deg$^2$ field of the LMC Bar. We used the
pixel light curves produced in Melchior et al
(1999\nocite{Melchior:1999a}, hereafter paper I) adopting looser
thresholds {\ch than} those used for the selection of microlensing
events in Melchior et al. (1998\nocite{Melchior:1998}, hereafter paper
II).  In Sect. \ref{sect:obs}, we summarise the characteristics of the
data set. In Sect. \ref{sect:sel}, we discuss the automatic selection
procedure used to keep all significant genuine variable stars.  In
Sect. \ref{sect:mag}, we introduce a magnitude estimate for each
variable and display its position in the colour-magnitude diagram. In
Sect. \ref{sect:ast}, we describe the procedure used to obtain the
equatorial coordinates. In Sect. \ref{sect:cross}, we cross-identify
the selected variable stars with existing databases.  Finally, we
provide the catalogue in Sect. \ref{sect:cat}.

\section{EROS CCD data and AGAPEROS pixel light curves}
\label{sect:obs}
\subsection{EROS CCD data}
We use the EROS-1 CCD dataset, taken at ESO over the period 1991
December 18 to 1992 April 11, using a 40~cm telescope with a wide
field camera composed of 16 CCD chips, each with 400$\times$579 pixels
of 1.21 arcsec (Arnaud et al., 1994b\nocite{Arnaud:1994b}, Queinnec,
1994\nocite{Aubourg:1995,Queinnec:1994}, Aubourg et al., 1995 and
Grison et al., 1995\nocite{Grison:1995}). Images of one field in the
LMC Bar were taken in two wide non-standard blue {\ch ($\bar{\lambda}
= 490$~nm)} and red {\ch ($\bar{\lambda} = 670$~nm)} filters, with a
mean seeing of 2.~arcsec.

The EROS-1 experiment (Arnaud et
al. 1994a,b\nocite{Arnaud:1994a,Arnaud:1994b}) was motivated by the
study of dark compact objects in the dark halo of our Galaxy and
contributed to show that the so-called brown dwarves could not be a
significant component of the dark matter (Renault et
al. 1997\nocite{Renault:1997}).

This dataset, treated in Paper~I, is composed of some 1\,000 images
per CCD and per colour spread over 120 days.  Only 10 CCD fields were
available in 91-92, so we restrict our analysis to this field of
0.25~deg$^2$.

\subsection{AGAPEROS pixel light curves}
{\ch For this selection, we use the pixel light curves produced in
Paper~I, as a first application to the EROS data of the Pixel Method
(Baillon et al. 1993)\nocite{Baillon:1993}.  In Paper~I, we described
in detail the data treatment we applied to the EROS-1 data set
described in the previous paragraph. The frames were first
geometrically and photometrically aligned with respect to a reference
image. In order to decrease the level of noise on these 8-12 min
exposures, we averaged the 10-20 frames available each night. Whereas
this increases our sensitivity to possibly dim LT$\&$LPV, {\em
this is not optimised for the detection of short period variable
stars, already studied elsewhere on the same data set (Grison et al
1995\nocite{Grison:1995}, Beaulieu et al. 1995\nocite{Beaulieu:1995b},
Hill $\&$ Beaulieu 1997\nocite{Hill:1997})}.

The data are then arranged in super-pixel light curves, and an empiric
seeing correction is then applied to each light curve.  The
limitations of this technique reside in the conversion of pixel fluxes
in magnitudes, which can be done efficiently with kernel convolution
techniques developed by To\-ma\-ney $\&$ Crotts (1996) and Alard
(1998)\nocite{Tomaney:1996,Alard:1998}. Due the large filters of the
EROS database and the short baseline (120 days) available with this
data set, we choose to postpone this step to the next paper, which
will produce periods with 3-years light curves together with a
cross-identification with the DENIS photometry.  In this paper, we
provide a magnitude estimate for one epoch, together with an
indication of blending, and put the emphasis on the selection
procedure and the cross-identification with other catalogues.

The light curves we use in the following to select variable stars
correspond to the super-pixel flux $\phi_{n} (t)$ measured at different
epochs $n$:
\begin{equation}
\phi_{n} (t) = \phi_{star}(t) + \phi_{\rm sky}
\end{equation}
where $\phi_{\rm sky}$ includes the sky and stellar background, and $
\phi_{star}(t)$ is the flux of the variable star.
}

\begin{table}
\caption[ ]{Selection procedure. Column (A) gives the parameters used
for thresholding, (B) gives the value of the threshold. (C) provides
the number of light curves kept at each step. Numbers surrounded by
boxes provide the actual threshold used and the actual number of
curves selected. Other numbers show how the weakening or strengthening
of these thresholds affect the number of selected light curves. This
procedure is further illustrated in Fig. \protect\ref{fig:eff}.}
\label{tab:data}
\begin{flushleft}
\begin{tabular}{lrr}
\noalign{\smallskip}
\hline
\noalign{\smallskip}
\multicolumn{1}{l}{Step   \hspace{0.1cm} Parameters} &
\multicolumn{1}{c}{Threshold} & \multicolumn{1}{c}{Number} \\ 
\multicolumn{1}{c}{(A)} & \multicolumn{1}{c}{(B)} &
\multicolumn{1}{c}{(C)}  \\\noalign{\smallskip} 
\hline\noalign{\smallskip}  
(0) Starting point after & excision &
\fbox{$2\,093\,584$ (100~\%)} 
\\\noalign{\smallskip} 
\hline\noalign{\smallskip}
 &{$< 0.7$} & \mbox{$133\,394$ (6.37~\%)} \\
(1) { Min($(\frac{\sigma_1}{\sigma_2})\vert_{B}$, 
$(\frac{\sigma_1}{\sigma_2})\vert_{R}$)} &\fbox{\bfseries $< 0.6$}
&\fbox{ $32\,067$ (1.53~\%)} \\
 &{$< 0.5$} & \mbox{$11\,369$ (0.54~\%)} \\\noalign{\smallskip} 
\hline\noalign{\smallskip}
 &{$> 300$} & \mbox{$24\,892$ (1.19~\%)} \\
(2) { Max($L_B$,$L_R$)} & \fbox{$>400$} &\fbox{
$20\,942$ (1.00~\%)} \\ 
 &{$> 500$} & \mbox{$17\,857$ (0.85~\%)}
\\\noalign{\smallskip}  
\hline\noalign{\smallskip}
(3) Clusters identification & &\fbox{$3\,782$
(0.18~\%)}  \\\noalign{\smallskip}   \hline\noalign{\smallskip}
(4)  {{ Max($L_B$,$L_R$)} } & \fbox{$>400$} &\fbox{
$3\,637$ (0.17~\%)} \\
  \hspace{0.5cm} { Min($(\frac{\sigma_1}{\sigma_2})\vert_{B}$, 
$(\frac{\sigma_1}{\sigma_2})\vert_{R}$)} &\fbox{\bfseries $< 0.6$}
&\fbox{ $3\,381$ (0.16~\%)} 
\\\noalign{\smallskip} \hline\noalign{\smallskip}
 &{$> 4$} & \mbox{$999$ (0.05~\%)} \\
(5) $N_{clus}$ &\fbox{$> 6$}
 &\fbox{ 747 (0.04~\%)} \\
&{$> 10$} &\mbox{544 (0.03~\%)} \\\noalign{\smallskip} 
\hline\noalign{\smallskip}
\end{tabular}
\end{flushleft}
\end{table}

\section{Selection of genuine variable stars}
\label{sect:sel}
\begin{figure*}
\resizebox{\hsize}{!}{\includegraphics{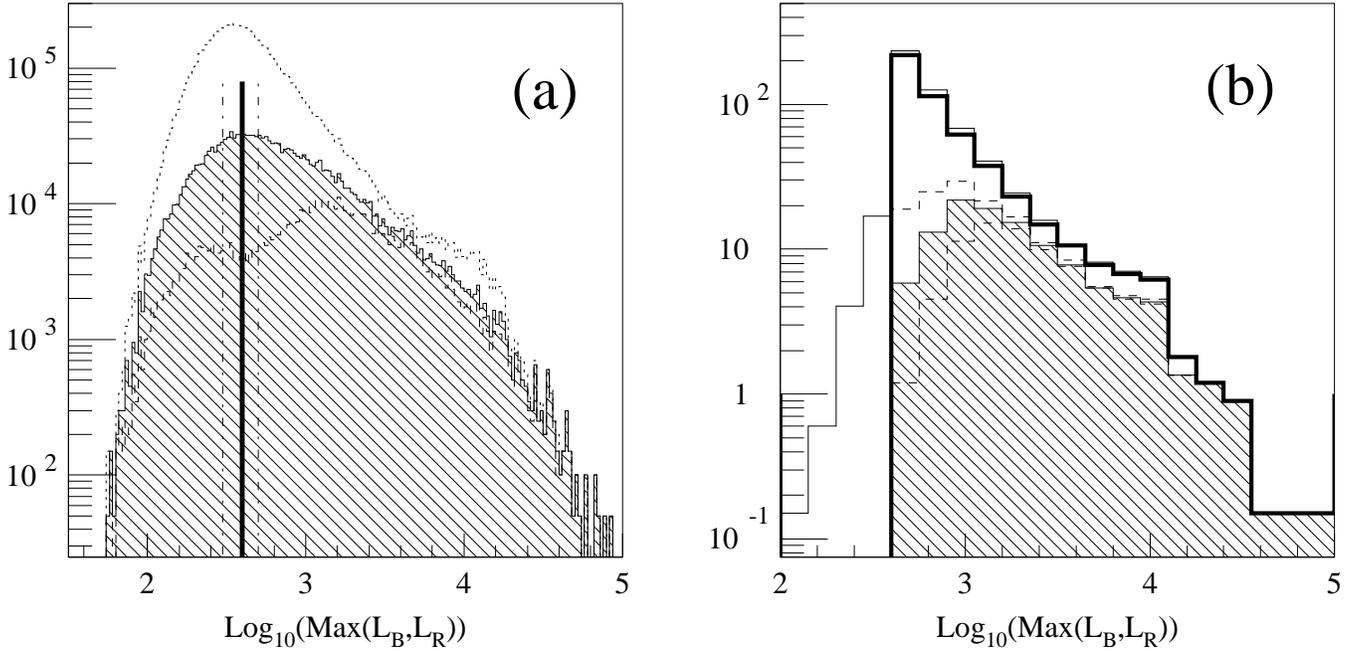}}
\caption{Effect of the selection procedure on the Max$(L_B,L_R)$
distribution. Panel {\bf a)} illustrates the effect of the steps (1)
and (2) of the selection procedure indicated in Tab.
\protect\ref{tab:data}. The hatched area corresponds to the pixels
kept at step (1). The dashed and dotted histograms each show how this
distribution is affected if this threshold is changed as indicated in
Tab.  \protect\ref{tab:data}. The thick line displays the effect of
step (2). The dot-dashed lines correspond to the
weakening/strengthening of this cut. Panel {\bf b)} illustrates the
effects of steps (4) and (5).  The large amplitude histogram shows the
distribution after step (3), the thick-line histogram shows the result
of step (4), and the hatched area corresponds to pixels kept at the
end of step (5). The dashed histograms show the sensitivity of this
last threshold.}
\label{fig:eff}
\end{figure*}
We apply an automatic selection of genuine luminosity variations to
the 91-92 dataset. Firstly, we introduce the definition of a
variation. Secondly, we adjust the thresholds in order to keep genuine
variations but reject most artifacts due to noise. Last, we carefully
inspect selected light curves close to bad pixels.

\subsection{Definition of a variation}
A baseline flux ($\phi_{bl}$) is calculated for each super-pixel light
curve by sorting all points in order of increasing flux, with
$\phi_{bl}$ being the 10$^{\rm th}$ sorted point. $\sigma_{bl}$ is the
error associated with the baseline flux determination.  For a sample
taken from a Gaussian distribution with a standard deviation $\sigma$
this estimate lies 1.3~$\sigma$ below the mean value of the
distribution.

Deviations from this baseline are recorded when measurements lie $3
\sigma_n$ above the baseline:
\begin{equation}
\sigma_n = \sqrt{{\sigma^{\prime}_{n}}^2 + {\sigma_{bl}}^2}
\end{equation}
where ${{\sigma}^{\prime}_{n}}$ is the error associated with each
super-pixel flux computed in Paper~I for night $n$.  These
deviations are quantified in each colour with a likelihood function
(L):
\begin{equation}
L = -\ln \left(\prod_{n \in \mathrm{bump}} P(\phi \ge \phi_n)\ \mbox{
given }\ \left\{
\begin{array}{c} \phi_{bl} \\ \sigma_n \end{array} \right. \right)
\label{lykelyhood}
\end{equation}
where $\phi_n$ is the super-pixel flux for the measurement $n$. All
the measurements above $\phi_{bl}$ are accounted for.

\subsection{Minimal threshold}
With the definition introduced above, we apply the selection
procedure summarised in Tab. \ref{tab:data}.

First of all we excise the pixels, covering 3.4\% of the CCD fields,
which exhibit obvious spurious variations (such as bad columns). In
addition, in order to remove automatically artifacts due to bad
pixels, we remove from the light curves the epochs for which there is
at least one pixel which datum is at zero within a 11$\times$11 window
centred on the super-pixel. (1) We require a regularity condition to
remove the noise: the ratio $\sigma_1/\sigma_2$ has to be smaller than
0.6 in at least one colour, with:
\begin{equation}
{\sigma_1}^2={\frac{2}{3(N-2)} {\sum_{n=2}^{N-1}}\left[ 
\frac{\phi_{n+1} - \phi_{n-1}}{2} - \phi_n \right]^2}
\end{equation}
\begin{equation}
{\sigma_2}^2={\frac{1}{(N-1)} {\sum_{n=1}^{N}}\left[ 
\phi_{n}  - \phi_{mean} \right]^2}
\end{equation}
where $\phi_{mean}$ is the mean super-pixel flux and $N$ is the total
number of measurements on each light curve.  (2) We then select the
light curves which vary such that $L>400$ in at least one colour.  As
there are as many super-pixels as pixels, a genuine variable is
expected to affect all the super-pixels within the seeing spot. (3) We
search for clusters of super-pixels (using a Friend of Friends
algorithm), and keep the central pixel of the clusters if (4) the
previous requirements are also satisfied for these pixels and, if (5)
the number $N_{clus}$ of super-pixels that compose each cluster is
larger than 6. These requirements eliminate most artifacts due to
bright stars\footnote{Super-pixels around a few bright stars exhibit
artifacts such that the spatial correlation of the position of these
super-pixels looks like a partial ring or an arc.}  and CCD defects
that do not exhibit a clear spatial PSF-like pattern. We finally keep
747 super-pixel light curves. The sensitivity of these thresholds is
illustrated in Fig. \ref{fig:eff}.  Among the 747 selected variations,
two have been counted twice\footnote{For computing reasons, each chip
has been cut into two pieces with an overlapping region through the
analysis.}, leaving 745 independent light curves.

\subsection{Artifact removals}
\label{sect:artif}
\begin{figure}
\resizebox{\hsize}{!}{\includegraphics{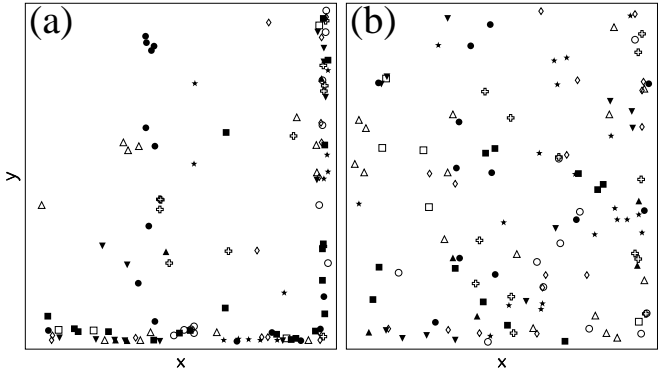}}
\caption{Artifact removals. Panel {\bf a)} displays the positions of
the 116 rejected light curves on the CCD chips. Panel {\bf b)} shows a
similar distribution but for those among the 237 light curves close to
CCD defects that have been kept.  Each symbol corresponds to a
different chip. Each panel corresponds to the size of the chips x$\in
[0,400]$, y$\in [0,579]$. }
\label{fig:cleand}
\end{figure}
We select 237 light curves for which there is at least one bad pixel
(saturated or set at zero) within a 21$\times$21 window centred on the
selected pixel for at least one epoch and at least one colour. A
careful visual inspection of these light curves shows 121 genuine
variable stars {\ch against} 116 artifacts, subsequently
removed. Figure \ref{fig:cleand} shows that the removed light curves
are mainly concentrated close to the edges, whereas the distribution
of the kept light curves (among the 237) is more uniform. We finally
end with a catalogue of 631 variable stars.

\section{Magnitude estimation}
\label{sect:mag}
\begin{figure}
\resizebox{\hsize}{!}{\includegraphics{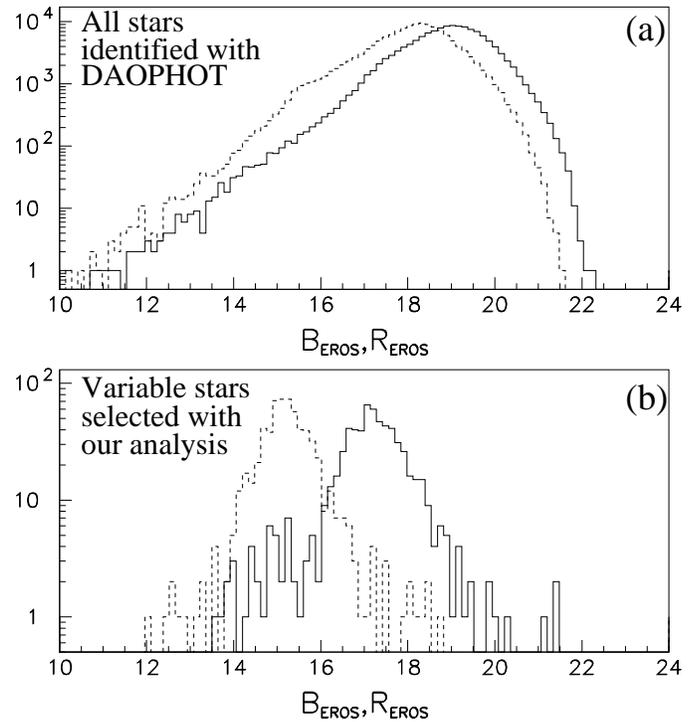}}
\caption{Luminosity functions in blue (full line) and in red (dashed
line) at the epoch JD=2448678.3. The upper histograms {\bf a)}
correspond to the magnitudes estimated for all the stars with
DAO\-PHOT. The lower histograms {\bf b)} correspond to the magnitudes
estimated as described in Sect. \ref{sect:pseudo} for the selected
variable stars.}
\label{fig:lf}
\end{figure}
Whereas the pixel method of analysis is able to detect variable stars
beyond the crowding limit, it does not measure photometry -- total
flux -- of these objects, that can be blended or even unresolved on
part of the light curves.  Obtaining their photometry would give a
first indication of the type of the variable stars.  Hence in this
section, we associate a magnitude and colour to each flux measurement.

\subsection{Pseudo-aperture photometry}
\label{sect:pseudo}
As discussed in Paper II, the flux of the super-pixel is composed of
the fraction of the flux of the star plus the {\ch local} background
(from sky and undetected stars). For our sample of variable stars, we
can presume that there is a star within the corresponding super-pixel
and that its flux significantly contributes to this super-pixel, at
least at the maximum of the variation. Because of the crowding
conditions, standard background estimates (circular annulus for
example, see Stetson (1987)\nocite{Stetson:1987}) fail and cannot be
used in an automatic way.  Hence, we choose to perform a
pseudo-aperture photometry as follows. For an image taken in the
middle of the period of observation (JD2448678.3) and with an average
seeing, we use the PSF fitting procedure of DAO\-PHOT (Stetson,
1987)\nocite{Stetson:1987} to measure the fluxes of the resolved
stars, and the background below them. This thus gives a local estimate
of the background that is the less affected by the crowding of the
field. Then for each selected super-pixel we look for the detected
star that is closest. The background estimate associated with this
star is supposed to be the same as the one present below the variable
star (and is even identical if the variable stars are resolved on this
reference frame). This background is subtracted from the super-pixel
flux. This flux is then corrected for the seeing fraction and
converted into a magnitude, corresponding to an isolated star{\ch.}

Whereas this definition of the sky-background is rather robust to the
crowding conditions, the magnitude estimation is not necessary so, as
some additional flux (stellar background) could contribute to the
super-pixel due to neighbours.  We thus quantify the blending with the
ratio $\phi_c / \phi_0$ computed as follows: the flux $\phi_c$ is the
averaged value computed along the light curve of the central pixel of
the super-pixel, and the flux $\phi_0$ is a similar average of the 8
surrounding pixels within the super-pixel. {\ch The behaviour of this
parameter is described in the Appendix.}

Optimised star detection with DAO\-PHOT allows to detect $135\,098$
stars in blue and $142\,870$ stars in red.  The corresponding
luminosity function computed with DAO\-PHOT for all the stars present on
the studied field, exhibited in Fig. \ref{fig:lf}a, shows that the
star detection is in first approximation complete down to magnitude 18
in red and 19 in blue. Figure \ref{fig:lf}b shows that the magnitude
distributions at the same epoch for the selected variable stars peak
at the bright end, whereas the stars are redder than average with
$B_{\rm EROS}-R_{\rm EROS} \simeq 2$. {\ch Whereas it is difficult to
compute our detection efficiencies as no reliable theoretical
distribution of variable stars is available, it is clear that we do
not detect a population of variable stars unresolved at minimum, even
though we do detect a tail of this distribution with very dim stars in
at least one colour.}

\subsection{Colour-magnitude diagram}
\label{sect:class}
\begin{figure*}
\resizebox{0.99\hsize}{!}{\includegraphics{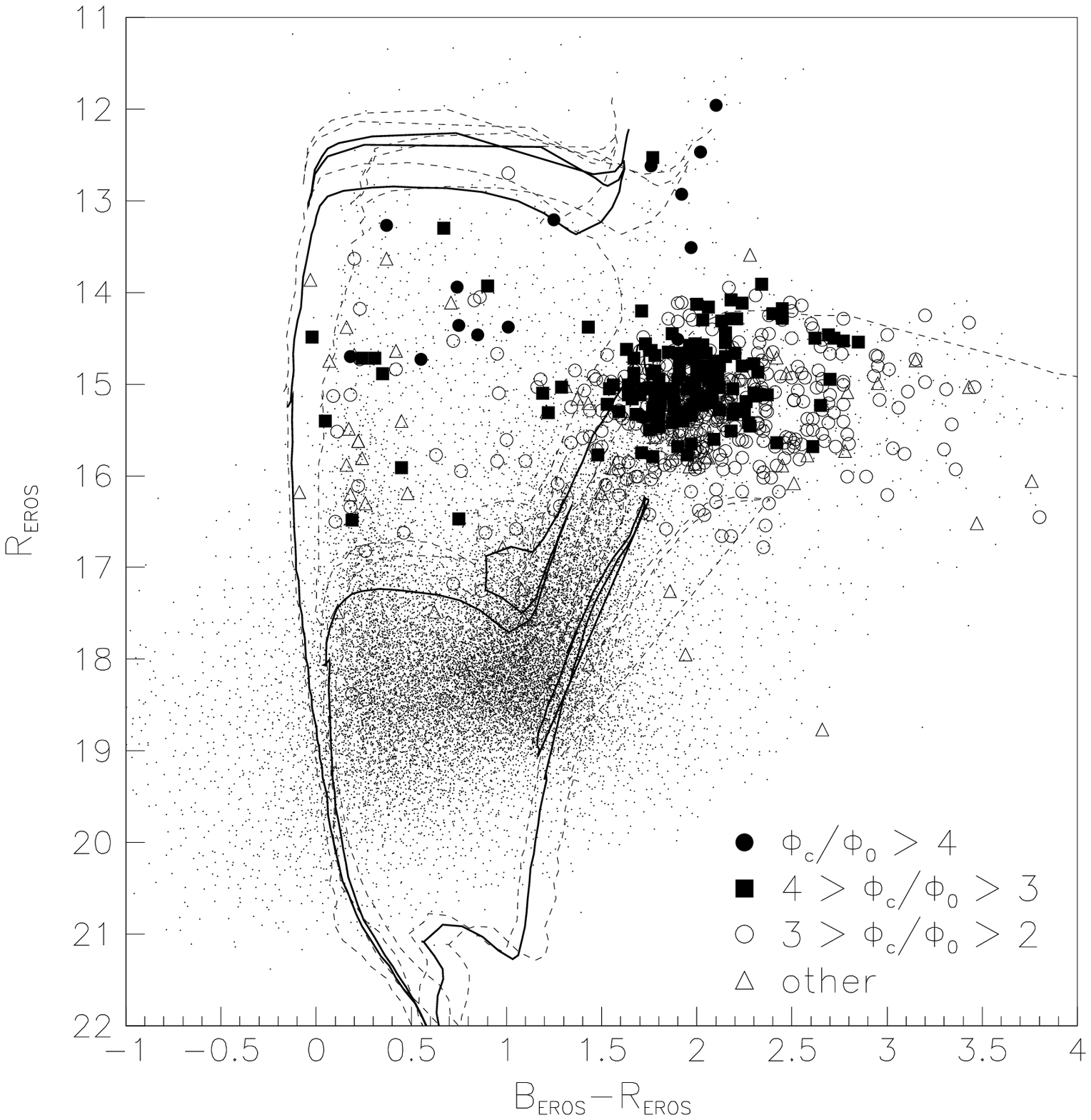}}
\caption{{\ch CMD} at JD$=$2448678.3: small dots
correspond to the stars detected with DAO\-PHOT{\ch,} symbols to the
631 selected variations. The different symbols correspond to different
ranges of the $\phi_c/\phi_0$ ratio {\ch: filled} symbols correspond
to stars with a high S/N ratio and not affected by crowding. {\ch The
superimposed isochrones} (full lines) are adapted from Bertelli et
al. (1994) {\ch to the EROS system (Grison et al. 1995) with a
foreground extinction $A_{\rm B_{\rm EROS}}=0.54$ and $A_{\rm R_{\rm
EROS}}=0.37$, deduced from reddening E(B-V)$=$0.15 measured by
Schwering $\&$ Israel (1991) with the extinction law from Cardelli et
al. (1989).} {\ch They} correspond to LMC metalicity ($Z=0.008$),
helium abundance ($Y=0.25$) with ages of $\log$(age)=7.4,8.4 and
9.4. The dashed lines show the uncertainties introduced on each
isochrone by the photometric transformation.}
\label{fig:colmag}
\end{figure*}
Figure \ref{fig:colmag} displays the position of the 631 variable
stars in the CMD. This CMD has been obtained with the resolved stars
detected with DAO\-PHOT in both colours. The conversion of the EROS
magnitudes into a standard system has been studied by Grison et
al. (1995)\nocite{Grison:1995}, but introduces significant systematic
errors especially for stars with $B_{EROS}-R_{EROS} > 1$. We hence
choose to work in the EROS system. The crowding limit prevents the
detection of stars with $R_{EROS}>19$. The main sequence and red clump
are clearly identified. Very little variations of the CMD are observed
from one chip to another suggesting a uniform stellar population
across this field.  The vast majority of the detected variable stars
lies in the red part. A visual inspection of the corresponding light
curves shows that they are compatible with {\ch LT$\&$LPV}. This
catalogue offers the perspective to systematically study the later
stages of stellar evolution within the LMC Bar. However, the
isochrones superimposed on Fig. \ref{fig:colmag} illustrate the
difficulty to classify variable stars with the sole CMD, as it is
impossible to disentangle their age and mass for a given
metalicity. The photometric system made it also inappropriate for
further interpretation with this sole data.  Due to their complexity
and the difficulties to model them, the {\ch LT$\&$LPV} are better
characterised in the IR, as addressed in a subsequent paper. Some
``bluer'' objects have also been detected and they will be classify
when we study their periodicity.

{\ch As shown in Fig. \ref{fig:colmag}, variable stars not
significantly affected by crowding ($\phi_c/\phi_0>3$ ) lie in areas of
the colour-magnitude diagram corresponding to stars expected
variable.} Those affected by blending and crowding will have to be
treated with caution. In the CMD areas where the larger number of {\ch
LT$\&$LPV} have been detected, in the magnitude ($R_{EROS}=14.6-15.8$)
and colour ($B_{EROS}-R_{EROS}=1.7-2.3$) ranges, about 17$\%$ of the
stars exhibit a variation detected with our analysis.

It is also clear that the vast majority of the detected variable stars
are above the crowding limit. The few outliers that can be noticed
correspond to stars unresolved in at least one colour, but their
number does not exceed 5~$\%$ of the total{\ch. In} terms of
microlensing, this means that events due to unresolved stars in the
LMC will not be significantly contaminated by the bulk of variable
stars. In further galaxies, like M\,31, variable stars will be a more
troublesome affair, and will have to be carefully studied (e.g. Crotts
\& Tomaney 1996\nocite{Crotts:1996}). However, high amplification
microlensing events are far less likely to be mimicked by an intrinsic
variation (Ansari et al. 1999)\nocite{Ansari:1999}, and will allow to
probe possible biases introduced by variable stars.

\section{Astrometric reduction}
\label{sect:ast}
The rectangular sky area covered by the EROS-1 chips is enclosed in a
circle of radius $\sim$30$\arcmin$ centred on the position $\alpha
=$05$^h$21$^m$52.1$^s$ ; $\delta=-69\degr34\arcmin16\arcsec$
(2000.0). In this region, the density of PMM astrometric
standards\footnote{http://ftp.nofs.navy.mil/projects/pmm/} is highly
irregular, e.g. four $\sim$20$\arcmin\times$ 15$\arcmin$ areas are
completely empty. This is mainly due to the crowding of the survey
plates used at {\ch USNO. We} therefore chose to use a U plate taken
at the ESO 1m Schmidt telescope to define the secondary astrometric
standards. This plate was scanned with the MAMA
microdensitometer\footnote{MAMA {(http://dsmama.obspm.fr)} is operated
by INSU (Institut National des Sciences de l'Univers) and Observatoire
de Paris.}, and reduced to the International Celestial Reference
System
(ICRS)\footnote{http://hpiers.obspm.fr/webiers/general/syframes/SY.html}
with the ACT
catalogue\footnote{http://vizier.u-strasbg.fr/cgi-bin/VizieR?-source=I/246}.
Thanks to the modest sensitivity of the U emulsion/filter combination,
the effects of crowding are reduced.  A remarkably regular
distribution of stars is detected with the SExtractor software (Bertin
$\&$ Arnouts, 1996)\nocite{Bertin:1996}, with on average 2\,700 stars
per chip.

{\ch The secondary standards ($\sim 1500$/chip) are identified to EROS
stars using a first linear transformation based on some 15 visually
cross-identified stars per chip. The astrometric reduction software
(Robichon et al. 1995\nocite{Robichon:1995}) is then run using this
set of references. After iteration, the final 2$^{\rm nd}$ order
reduction keeps 1000 stars per chip with final rms deviation of the
order of 0.2$\arcsec$.

According to the galactic
model\footnote{http://WWW.obs$-$besancon.fr/www/modele/s{${}_{-}$}tab.html}
from Besan\c con, more than 95$\%$ of the secondary references belong
to LMC. Furthermore, we check that the global proper motion of the LMC
(Kroupa $\&$ Bastian, 1997\nocite{Kroupa:1997}), together with upper
limits on its internal velocity dispersion cannot affect the positions
of our standards by more than 30~mas, over the 848-days separating the
ESO plate from the EROS reference.  }

\section{Cross-identifications} 
\label{sect:cross}
{\ch The detected variable stars are uniformly distributed along the
CCD chips, and no obvious bias (due e.g. to internal extinction) are
suspected.}  In order to appreciate the sensitivity of our analysis,
we cross-identified this catalogue with previous works.  Among the 631
variable stars presented here, {\ch 72} have already been studied
elsewhere. We considered the main catalogues which covered this area,
namely the catalogues of {\ch LT$\&$LPV} by Hughes
(1989)\nocite{Hughes:1989}, eclipsing binaries and cepheids by EROS
(Grison et al., 1994 and Beaulieu et al.\
1996)\nocite{Grison:1994,Beaulieu:1996}, radio sources (Marx et al.,
1999), X-ray sources (Haberl \& Pietsch, 1999)\nocite{Haberl:1999}, as
well as the General Catalogue of Variable Stars\footnote{We used the
electronic version of the GCVS
(http://www.sai.msu.su/groups/cluster/gcvs/gcvs/), supported by the
Russian Foundation for Basic Research.} (Artiukhina et al.,
1995)\nocite{Artiukhina:1995} and the NED\footnote{The NASA/IPAC
Extragalactic Database (NED) is operated by the Jet Propulsion
Laboratory, California Institute of Technology, under contract with
the National Aeronautics and Space Administration.} data-base.  We
define a genuine cross-identification any object whose offsets with
respect to the variable objects detected here are smaller than $\Delta
\alpha_{\rm max}=5\arcsec$ and $\Delta \delta_{\rm max}=5\arcsec$.
\begin{figure}
\resizebox{\hsize}{!}{\includegraphics{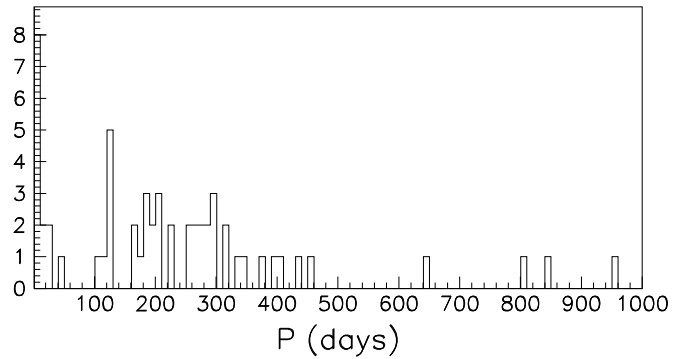}}
\caption{Histogram of the periods. We present here the periods
measured elsewhere which correspond to 57 of the 631 variable stars
detected here.}
\label{fig:per}
\end{figure}

Among the 58 {\ch LT$\&$LPV} found by Hughes
(1989)\nocite{Hughes:1989} in this field, {\ch 41 have been detected
here, 12 are missed due to CCD defects (one of those is rejected with
the visual inspection described in Sect. \ref{sect:artif}).} {\ch This
gives the order of magnitude of the completeness of our sample over a
120-day window: we typically miss up to 20$\%$ of genuine variable
stars due to CCD defects. More interestingly, the last 5 missing {\ch
LT$\&$LPV} escape our selection for no obvious reason. One of them
(SHV0527122-695006) has been subsequently observed in the
near-infrared (Hughes $\&$ Wood 1990)\nocite{Hughes:1990}: it did not
follow the (K, $\log P$) relation, and was interpreted as a supergiant
or a foreground Mira. According to the DAO\-PHOT magnitudes measured
at epoch JD=2448678.3 B$_{\rm EROS}=15.7$ and R$_{\rm EROS}=13.7$, it
is consistent with a supergiant.  The 4 remaining, namely
SHV0516251-693241, SHV0519415-693441, SHV0522220-694441,
SHV0522251-692902 -- not observed subsequently by Hughes $\&$ Wood
(1990) -- were reported by Hughes (1989) with a low amplitude ($\Delta
I \simeq 0.5$~mag.) and 3 of them were only marginally
periodic. Moreover, our non-detection can be explained for some of
them by changes in mass-loss rates as suggested by Whitelock (1997).
The extension of this work with a longer baseline together with DENIS
photometry is expected to provide further arguments about this kind of
behaviour (in preparation).}

The short periodic variable stars from the EROS catalogue (Grison et
al., 1994, and Beaulieu et al., 1996) are missed here for most of
them: short timescale variations are broken by the averaging procedure
as explained in Sect. \ref{sect:obs}. Only 22 out of the 181 variable
stars previously detected by EROS are present in this catalogue. In
addition, 5 of the 7~pre-main-sequence candidates published by
Beaulieu et al. (1996)\nocite{Beaulieu:1996} have been detected.  None
of the radio-sources, detected by Marx et
al. (1997)\nocite{Marx:1997}, are lying in the studied field. 2 X-ray
sources detected with ROSAT (Haberl \& Pietsch,
1999)\nocite{Haberl:1999} are present in the field, but {\ch none of
our variable stars lies in the 90$\%$CL error box of those
sources}. 29 extragalactic sources (but not necessarily variable) from
NED are present in the field, none is among the variable stars
detected here. Last, whereas 129 sources from the GCVS are lying in
the field, 52 are selected here. Most of them were also in the
catalogues mentioned above. According to the GCVS classification,
among these 52 variables, 17 are Miras, 26 semi-regulars, 2
irregulars, and 7 cepheids.

A total of 72 sources out of 631 were previously reported, 57 of which
have a published period. Those are presented in Fig. \ref{fig:per}:
they are clearly of the order of 100-300 days. A visual inspection
confirms that besides a few short periods, these timescales are quite
representative of the bulk of a distribution dominated by {\ch
LT$\&$LPV}.
\begin{figure}
\resizebox{\hsize}{!}{\includegraphics{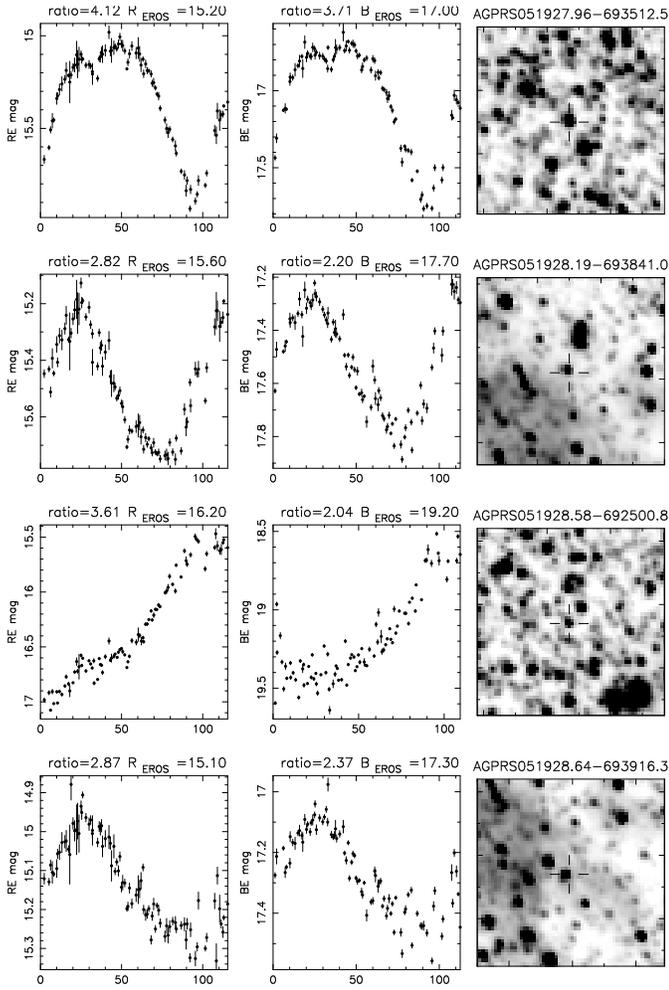}}
\caption{Examples of red variables. The light curves in red and blue
together with the corresponding finding chart in red are displayed for
4 red variable stars from the catalogue. The $\phi_c/\phi_0$ ratio is
provided for each light curve, as well as the magnitude at
JD$=$2448678.3. Note the fact that the large dispersion with respect
to the errors observed for some blue light curves is due to blending
and those light curves have a low \protect$\phi_c/\phi_0$ ratio.}
\label{fig:ex1}
\end{figure}
\begin{figure}
\resizebox{\hsize}{!}{\includegraphics{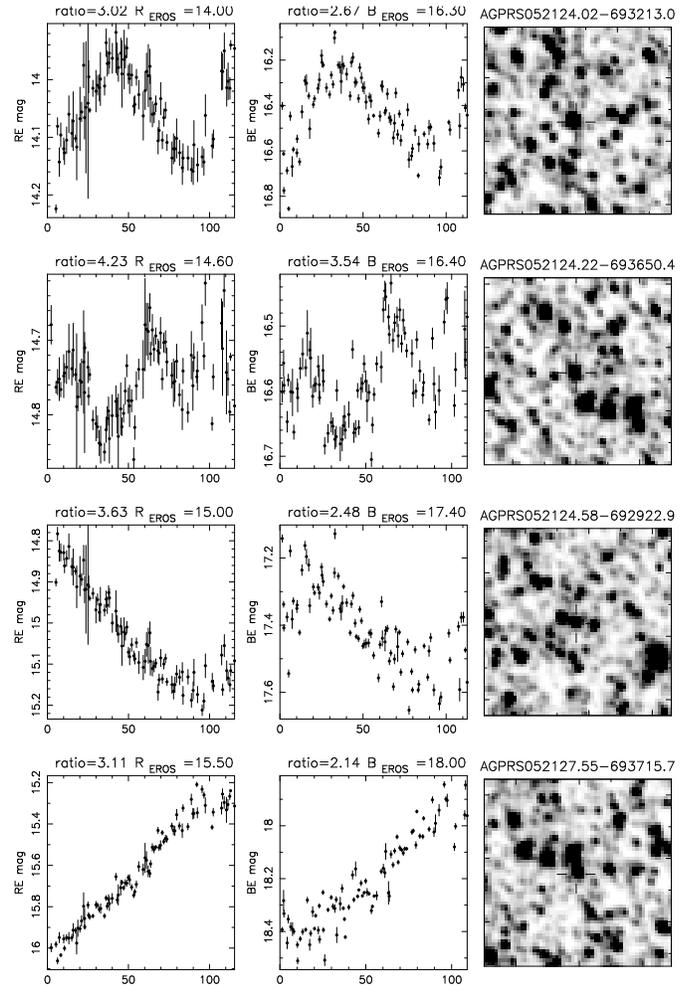}}
\caption{Examples of red variables. Same as Fig. \ref{fig:ex1}.}
\label{fig:ex2}
\end{figure}

\section{Catalogue}
\label{sect:cat}
{\ch LT$\&$LPV} (``Red'' variable stars) represent the bulk of the
stars presented in this catalogue.  Figures \ref{fig:ex1} and
\ref{fig:ex2} show typical light curves detected with the
corresponding finding charts.  Amplitudes measured over 120 days can
be as low as 0.1~mag up to several magnitudes, depending on the
timescale, and brightness range. The limitations (in the photometry)
due to the crowding are also illustrated. The error bars (estimated in
Paper~I) account for the level of photon noise and other residual
errors, but they do not account for the effect of crowding on the star
flux. For light curves with $\phi_c/\phi_0\le 3$, error bars are
underestimated with respect to the scatter along the variation. This
explains why the amplitude of variations in blue can be smaller (or
comparable) than in red for a few {\ch LT$\&$LPV}: the crowding is
more important in blue and larger amplitudes of variation are expected
on dimmer stars. Hence, we do not attempt to extract information about
the amplitude, but provide magnitude estimates at one epoch with the
$\phi_c/\phi_0$ ratio. For most of them, the timescale of the
variation is much longer than the period of observation, and the
periodicity will not be studied here.

Table~2 contains the following informations. {\it
Column}~1: The position, at J2000.0 epoch, is incorporated into the
name of each variable, in accordance with IAU convention (Dickel et
al. 1987)\nocite{Dickel:1987}; {\it Column}~2: N/X/Y: chip number N
and position (X,Y) on the chip. There were 10 chips on the EROS-1 CCD
camera. {\it Column}~3: the $\phi_c/\phi_0$ ratio in blue and red;
{\it Column}~4: the $B_{EROS}$ and $R_{EROS}$ magnitude estimate at
epoch JD=2448678.3; {\it Column}~5: the number of stars detected by
DAO in the super-pixel at epoch JD=2448678.3 in B/R; {\it Column}~6
and~7: the $B_{EROS}$ and $R_{EROS}$ magnitudes (at the same epoch) of
the closest stars as estimated by DAO\-PHOT. Figures in bracket (on a
second line) provide the magnitude of another star detected in the
super-pixel.  {\it Column}~8: identification with the other databases,
discussed in the text. The Harvard variable numbers are given, as well
as variables (WBP) detected by Wood et al. (1985)\nocite{Wood:1985}.

The finding charts in red are presented for each variable in
Fig.~8.  They correspond to the epoch for which the
stellar magnitude and blending has been studied (JD=\-2448678.3). Each
chart is labelled with the name of the variable. A cross indicates the
position at which the variable star has been detected.

\section{Conclusion}
Besides being a by-product of the microlensing searches, the variable
stars constitute their background and their understanding and
discrimination are important (e.g. della Valle \& Livio
(1996)\nocite{dellaValle:1996}). They need to be catalogued for any
on-line microlensing searches (cf. MACHO and EROSII). In this paper,
we have produced a comprehensive catalogue of 631 variable stars
selected over a 0.25 deg$^2$ field in the LMC Bar with the Pixel
Method applied to a 120-day window, with a sampling of about one
measurement per day. For each variable star, we provide an astrometric
position accurate within $~1\arcsec$, together with a finding chart.
The study of their position in the CMD shows that this catalogue is
dominated by a population of {\ch LT$\&$LPV}, while a few ``bluer''
variables have also been detected.  Cross-identification with existing
catalogues shows that 11~\% of them have already been studied
elsewhere. We hence enlarge by a factor of 10 the number of {\ch
LT$\&$LPV} detected in this area of the LMC Bar.

The pixel analysis presented here allows a selection of variable
objects independently of a photometry. It hence uses all the
information present in the frames but does not provide a photometry
for these objects. It is complementary to the image subtraction
techniques. We have shown that not surprisingly the pixel light curves
are significantly polluted by blending, especially in blue and that
their conversion in magnitude can only be rough. Hence, for each
variable star, we provide an indicator of blending in both colours.
The preliminary study of these variable stars together with their
cross-identification with previous works show that our photometry does
not introduce significant bias on the overall distribution. Recent
improvements of the image subtraction techniques developed for
microlensing searches (Alard 1999, Alcock et al.\
1999a,b)\nocite{Alard:1999,Alcock:1999a,Alcock:1999b} will have to be
considered for the future to improve the photometry of the variable
stars detected in such crowded fields (e.g. Olech et
al. 1999\nocite{Olech:1999}). However, most of these objects are Long
Period Variables better characterized with IR photometry than optical
non-standard photometry anyway.

The limitation of this dataset is hence reached and it needs to be
complemented by other data.  In a companion paper, we will extend
these selected light curves to the whole EROS~I baseline (850 days
window), with the 1992-94 data taken with a different set of
filters. This will enable us to study the periodicity of these
variable stars. In a forthcoming paper, we will cross-identify these
variable stars with the IJK photometry of the DENIS catalogue.

\begin{flushleft}
{\bf Table~2. Catalogue of the variable objects detected in this
 paper. They are sorted in order of increasing right ascension.
It can be retrieved at
http://cdsweb.u-strasbg.fr/htbin/Cat?J/A$\%$2bAS/145/11.}
\end{flushleft}
\begin{flushleft}
{\bf Fig.8. $1.2\arcmin \times 1.2\arcmin$ finding charts for each
variable in the R$_{\rm EROS}$ filter. It is available in the paper
version of the article at A$\&$AS or electronically at
http://www.edpsciences.com/articles/astro/full\-/2000/13/ds9463/node10.html.
}
\end{flushleft}

\acknowledgements{We thank the EROS collaboration
(http:\-//www.lal.in2p3.fr/\-recherche/\-eros/\-erosa.html) for giving
us their 91-92 data, and Marc Moniez for a careful reading of this
manuscript.  The U plate used for the astrometry was taken at the ESO
Schmidt telescope at La Silla. We thank the MAMA team for scanning and
reducing this plate. We acknowledge Dave Monet for the informations he
kindly provided on the PMM/USNO catalogue.  We are particularly
grateful to C. Lamy for her useful help on data handling during this
work, done in part on the computers of PCC-Coll\`ege de France.
A.-L. Melchior has been supported by a European contract
ERBFMBICT972375 at QMW.  This research has made use of the NASA/IPAC
Extragalactic Database (NED) which is operated by the Jet Propulsion
Laboratory, California Institute of Technology, under contract with
the National Aeronautics and Space Administration. The Vizier service
(Ochsenbein, 1997)\nocite{Ochsenbein:1997} at CDS has been used
throughout this work.}


\newpage
\appendix
\section{Limitations of the Pixel Method}
\begin{figure}
\resizebox{\hsize}{!}{\includegraphics{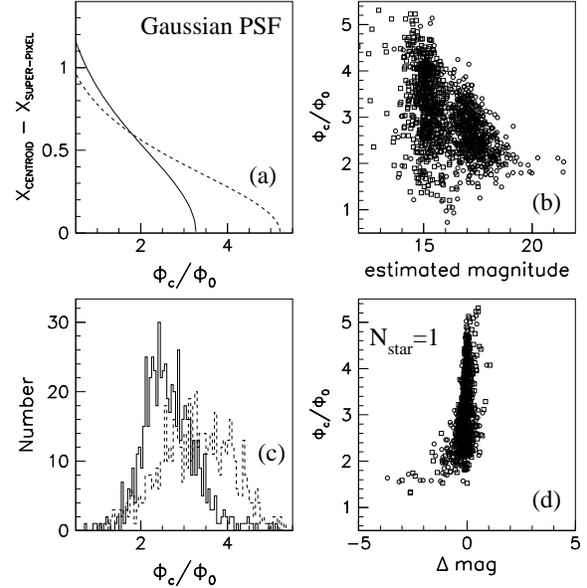}}
\caption{Quantification of blending.  Square symbols correspond to red
measurements and circle to blue.  {\bf a)} The position of the star's
centroid as a function of the $\phi_c/\phi_0$ ratio is displayed for a
single Gaussian PSF (seeing=2.0-1.6 arcsec), without any crowding. In
addition, cylindrical symmetry is assumed.  {\bf b)} The measured
ratio $\phi_c/\phi_0$, which also includes a measure of the S/N ratio,
shows a strong correlation with the magnitude of the stars at
JD=2448678.3.  {\bf c)} Histograms of the measured $\phi_c/\phi_0$
ratio in blue (full line) and in red (dashed line) of all the selected
light curves: blending is stronger in blue {\bf d)} Dependence of the
ratio $\phi_c/\phi_0$ as a function of the difference between the
estimated magnitude and the magnitude of the closest star detected by
DAO\-PHOT at the epoch JD=2448678.3 for $N_{star} = 1$, discussed above
and in Fig. \ref{fig:magcomp}b.}
\label{fig:ratio}
\end{figure}
\begin{figure}
\resizebox{\hsize}{!}{\includegraphics{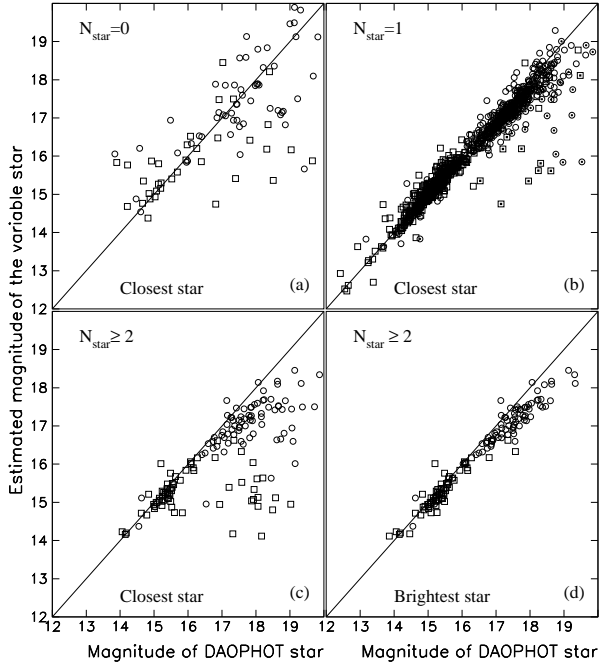}}
\caption{{\ch Study of the crowding conditions of each variable star
of the catalogue with} magnitude estimates at the epoch
JD=2448678.3. The squares correspond to the red measurements and the
circles to the blue ones. Each panel corresponds to a given value of
$N_{star}$, {\ch the number of stars in the super-pixel, as detected
by DAO\-PHOT}. {\bf a)} These points gather stars unresolved at the
epoch studied (the majority in blue), stars missed by DAO\-PHOT (bad
PSF or other defect), mis-alignment of the super-pixel with respect to
the variable star, and few apparently spurious variations.  No
correlation is observed with the closest star. {\bf b)} There is a
good correlation between the magnitude of the star detected by
DAO\-PHOT and the estimated magnitude, except for a few ones. Filled
symbols are measurements for which there is a brighter star within
3~pixels of the centre of the super-pixel, and are clearly affected by
blending. The observed dispersion, and in particular the points for
which the DAO\-PHOT magnitude is brighter, is mainly due to the fact
that the precision on the position of the centre of the super-pixel is
1 pixel, whereas the position of the centroid of the star defined by
DAO\-PHOT is much more accurate.  Panel {\bf c)} displays the
correspondence between the estimated magnitude and the magnitude of
the {\em closest} star detected by DAO\-PHOT. A significant component
of blending can be observed. Panel {\bf d)} shows a better correlation
as the {\em brightest} star detected by DAO\-PHOT in the super-pixel
has been used.  At the dim end, the crowding limit produces
unavoidable blending.}
\label{fig:magcomp}
\end{figure}

Figure \ref{fig:ratio} displays the behaviour of the ratio $\phi_c /
\phi_0$ computed for the variable stars of the catalogue.
Figure \ref{fig:magcomp} compares our magnitude estimate with the
DAO\-PHOT one for the selected light curves. 

\end{document}